\documentclass[fleqn,12pt,twoside]{article}
\usepackage{amsmath,espcrc1,graphicx,epsfig,wrapfig}

\begin{document}
\pagestyle{plain}

\title{Comparison of numerical solutions for $Q^2$ evolution equations}
\author{S. Kumano and T.-H. Nagai
\address{Department of Physics, Saga University,
              Saga, 840-8502, Japan}
\thanks{kumanos@cc.saga-u.ac.jp, 03sm27@edu.cc.saga-u.ac.jp,
        http://hs.phys.saga-u.ac.jp}}
\maketitle

\vspace{-4.5cm}
\hfill{SAGA-HE-199-04}
\vspace{+3.9cm}

\begin{abstract}
$Q^2$ evolution equations are important not only for describing hadron
reactions in accelerator experiments but also for investigating
ultrahigh-energy cosmic rays. The standard ones are called DGLAP
evolution equations, which are integrodifferential equations.
There are methods for solving the $Q^2$ evolution equations for
parton-distribution and fragmentation functions. Because the equations
cannot be solved analytically, various methods have been developed for
the numerical solution. We compare brute-force, Laguerre-polynomial, and
Mellin-transformation methods particularly by focusing on
the numerical accuracy and computational efficiency. An efficient
solution could be used, for example, in the studies of a top-down
scenario for the ultrahigh-energy cosmic rays. 
\vspace{1pc}
\end{abstract}

\section{Introduction}
\label{intro}

High-energy hadron reactions are described in terms of parton-distribution
functions (PDFs) and fragmentation functions (FFs). There are
parametrizations for the PDFs \cite{pdfs} and FFs \cite{ffs}.
Using these functions, cross sections of high-energy hadron reactions are
evaluated. Precise calculations of these cross sections are important for
finding any new physics beyond the current theoretical framework. 

The PDFs depend on two kinematical variables $x$ and $Q^2$.
They are defined by $Q^2=-q^2$ and $x=Q^2/(2p\cdot q)$ in lepton
scattering with the momentum transfer $q$ and the hadron momentum $p$.
The FFs depend on $Q^2$ and another variable $x=2E_h/\sqrt{s}$, where
$E_h$ is the hadron energy and $\sqrt{s}$ is the center-of-mass energy.
Their $Q^2$ dependence is called scaling violation, which is calculated
by the DGLAP (Dokshitzer-Gribov-Lipatov-Altarelli-Parisi) evolution
equations \cite{dglap} in the perturbative QCD region.

The $Q^2$ evolution equations are frequently used in describing high-energy
hadron reactions. Because the PDFs and FFs vary significantly in the current
accelerator-reaction range, $Q^2$=1 GeV$^2$ to 10$^5$ GeV$^2$, the $Q^2$
dependence should be calculated accurately. Furthermore, it is known that
high-energy cosmic rays have energies much more than the TeV scale.
Analytical forms of current PDFs and FFs are supplied typically
in the GeV region, so that they have to be evolved to the scale
which could be more than TeV in order to use them for investigating
the cosmic rays \cite{toldra,topdown}.

A useful evolution code was developed in Ref. \cite{toldra} for
the cosmic ray studies. The Laguerre-polynomial method was used for
solving the evolution equations. Splitting functions, PDFs, and FFs are
expanded in terms of the Laguerre polynomials, and then the evolution is
described by a simple summation of their expansion coefficients.
In fact, the method is very efficient for solving the equations
in comparison, for example, with a direct integration method
\cite{bf,bfsaga}. However, because the Laguerre polynomials $L_n(-\ln x)$
go to infinity in the limit $x\rightarrow 0$, it could have an accuracy problem
in the small $x$ region where high-energy reactions are sensitive.
In this paper, we discuss evolution results by the Laguerre method
\cite{toldra,lag} in comparison with the ones by other solution methods,
``brute-force" \cite{bf} and Mellin-transformation \cite{mellin} methods.
In particular, evolution accuracy and computation time are compared.
It is the purpose of this paper to clarify the advantages and disadvantages
of these numerical solution methods for a better description of
high-energy hadron reactions including the high-energy cosmic rays.
In particular, the FF evolution could be used for studying a top-down
scenario in order to determine the origin of ultrahigh-energy cosmic
rays, namely from a decay of a superheavy particle \cite{topdown}.

This paper consists of the following. The DGLAP evolution equations
are introduced in Sec.\,\ref{q2evol}, and numerical solution methods
are explained in Sec.\,\ref{methods}. Evolution results and their
comparisons are discussed in Sec.\,\ref{results}. The results are
summarized in Sec.\,\ref{summary}. 

\section{$Q^2$ evolution equations}
\label{q2evol}

From the cross-section measurements of high-energy lepton-hadron,
hadron-hadron, and lepton-annihilation reactions,
the PDFs and FFs are extracted. The PDFs and FFs are expressed
in terms of the two kinematical variables $x$ and $Q^2$. 
A PDF or a FF is expressed $f(x,Q^2)$ in the following.
We investigate the standard $Q^2$ evolution equations, which are called
the DGLAP evolution equations \cite{dglap}. The flavor nonsinglet equation
is written as
\begin{equation}
\frac{\partial}{\partial\ln Q^2} \ f_{_{NS}}(x,Q^2) = 
\frac{\alpha_s (Q^2)}{2\pi} \, \int_x^1 \frac{dy}{y} \,
P_{_{NS}}(x/y) \,
f_{_{NS}}(y,Q^2) ,
\label{eqn:dglap1}
\end{equation}
where $f_{_{NS}}(x,Q^2)$ is a nonsinglet (NS) function,
$P_{_{NS}}(x)$ is a nonsinglet splitting function,
and $\alpha_s(Q^2)$ is the running coupling constant.
The splitting functions for parton distributions and fragmentation functions
are identical in the leading order (LO) of $\alpha_s$; however, they differ
if higher order corrections are included \cite{ffnlo}.
In order to make the evolution equation slightly simpler, the variable $t$
is used instead of $Q^2$:
\begin{equation}
t \equiv - \frac{2}{\beta_0} \ln 
      \biggl [ \frac{\alpha_s(Q^2)}{\alpha_s(Q_0^2)} \biggr ] , 
\label{eqn:t}   
\end{equation}
where the running coupling constant in the leading order (LO) is given by
$ \alpha_s(Q^2) = 4\pi/[\beta_0 \ln (Q^2/\Lambda^2)] $
with the QCD scale parameter $\Lambda$. 
The constant $\beta_0$ is expressed as
$\beta_0= 11 C_G/3 - 4 T_R N_f /3$ with $C_G=N_c$ and $T_R=1/2$.
Here, $N_c$ is the number of color ($N_c$=3) and  $N_f$ is the number
of flavor. The $Q_0^2$ in Eq. (\ref{eqn:t}) indicates the initial $Q^2$
where the evolved function is provided.
Using this variable $t$ in Eq. (\ref{eqn:dglap1}), we obtain 
\begin{equation}
\frac{\partial}{\partial t} \, f_{_{NS}}(x,t) 
           = \int_x^1\frac{dy}{y} \, P_{_{NS}}(x/y) \, f_{_{NS}}(y,t) ,
\label{eqn:ns}
\end{equation}
for the nonsinglet evolution.

In the flavor singlet case, the evolution is described by
two coupled integrodifferential equations:
\begin{equation}
\frac{\partial}{\partial t} \, {\bf f}(x,t)
=\int_x^1\,\frac{dy}{y} \, {\bf P}(x/y) \, {\bf f}(y,t) ,
\label{eqn:sing}
\end{equation}
where the matrices ${\bf f}$ and ${\bf P}$ are defined by
\begin{equation}
{\bf f}(x,t) = 
 \begin{pmatrix}
    f_{_S}(x,t) \\
    f_{g}(x,t)
 \end{pmatrix} , \ \ \ \ 
{\bf P}(x)=
 \begin{pmatrix}
   P_{qq}(x) & 2 N_f P_{ij}(x) \\
   P_{ji}(x) &       P_{gg}(x)
 \end{pmatrix} .
\end{equation}
The indices $i$ and $j$ indicate $ij=qg$ for the PDFs
and $ij=gq$ for the FFs.
The functions $P_{qq}$, $P_{qg}$, $P_{gq}$, and $P_{gg}$
are splitting functions. The function $P_{ij}$ determines
the probability of the splitting process 
that a parton $j$ with the momentum fraction $y$ 
splits into a parton $i$ with the momentum fraction $x$
and another parton and the $j$-parton momentum is reduced
by the fraction $z$. 
In the LO, the splitting functions are expressed as
\begin{align}
P_{qq}(x) =& P_{_{NS}}(x) =
             C_F \, \left[\, \frac{1+x^2}{(1-x)_+}
                     + \frac{3}{2} \, \delta (1-x)\, \right] , 
\nonumber \\
P_{qg}(x) =& T_R \left [\, x^2 + (1-x)^2\ \right] ,
\nonumber \\
P_{gq}(x) =& C_F \ \frac{1+(1-x)^2}{x}  ,
\\
P_{gg}(x) =& 2 \, C_G \left[ \, \frac{x}{(1-x)_+}  + 
                         \frac{1-x}{x} \ + \ x(1-x)  + 
            \left( \frac{11}{12} - \frac{1}{3}\frac{N_fT_R}{C_G}\right)
            \, \delta (1-x) \, \right] ,
\nonumber
\end{align}
where $C_F$ is given by $C_F=(N_c^2-1)/(2N_c)$ and
$1/(1-x)_+$ is defined by
$\int_0^1 dx \, g(x)/(1-x)_+ = \int_0^1 dx \, [g(x)-g(1)]/(1-x)$
with an arbitrary function $g(x)$.

We need to solve the nonsinglet and singlet evolution equations
in Eqs. (\ref{eqn:ns}) and (\ref{eqn:sing}). These are not simple
integrodifferential equations, so that an efficient numerical 
method should be investigated. In the next section,
three popular numerical methods are explained.

\section{Numerical Methods for solving $Q^2$ evolution equations}
\label{methods}

There are various ways for solving the DGLAP equations. 
In this section, we explain three popular methods, brute-force,
Laguerre-polynomial, and Mellin-transformation methods.
In the following subsections, only the nonsinglet evolution is explained
because the singlet evolution can be solved in the same way.

\subsection{Brute-force method}
\label{bf}

The simplest way is possibly to use the brute-force method
\cite{bf,bfsaga}. It may seem to be too simple, but it is especially suitable
for solving more complicated equations with higher-twist terms \cite{mq}.
These equations could not be easily handled by the orthogonal-polynomial
methods such as the Laguerre-polynomial one in Sec.\,\ref{laguerre} and by
the Mellin-transformation method in Sec.\,\ref{mellin}. Furthermore,
a computer code is so simple that the possibility of a program mistake
is small, which means the code could be used for checking other
numerical methods. These are the reasons why it was investigated in
Ref. \cite{bfsaga}. 

In the brute-force method, the two variables $t$ and $x$ are divided into
small steps, and then the differentiation and integration are defined by
\begin{equation}
\frac{\partial f(x,t)}{\partial t}\Rightarrow
\frac{f(x_i,t_{j+1})-f(x_i,t_j)}{\Delta t_j} ,
\ \ \ \ 
\int dx f(x,t)\Rightarrow \sum_{k=1}^{N_x}\Delta x_k f(x_k,t_j)
,
\label{eqn:bfint}
\end{equation}
where $\Delta t_j$ and $\Delta x_k$ are the steps at the positions
$j$ and $k$, and they are given by $\Delta t_j=t_{j+1}-t_j$ and
$\Delta x_k=x_k-x_{k-1}$. The numbers of $t$ and $x$ steps are
denoted $N_t$ and $N_x$, respectively.
Applying these equations to Eq. (\ref{eqn:ns}), we write the nonsinglet
evolution from $t_j$ to $t_{j+1}$ as
\begin{equation}
q_{_{NS}}(x_i, t_{j+1})  =q_{_{NS}}(x_i,t_j)
+\Delta t_j\sum_{k=i}^{N_x}
\frac{\Delta x_k}{x_k}P_{qq}(x_i /x_k) q_{_{NS}}(x_k,t_j)
.
\label{eqn:bfns}
\end{equation}
If the distribution $q_{_{NS}}$ is supplied at $t_1=0$, the next one
$q_{_{NS}}(x, t_{2})$ can be calculated by the above equation.
Repeating this step $N_t-1$ times, we obtain the final distribution
at $t_{N_t}$. However, it is obvious that the step numbers $N_t$
and $N_x$ should be large enough to obtain an accurate evolution
result.

\subsection{Laguerre polynomial method}
\label{laguerre}

The evolution equations could be solved by expanding the distribution 
and splitting functions in terms of orthogonal polynomials. A popular
method of this type is to use the Laguerre polynomials \cite{toldra,lag}.
They are defined in the region from 0 to $\infty$, so that
the variable $x$ should be transformed to $x'$ by the relation
$x'=-\ln \, x$. 

The nonsinglet evolution is discussed in the following.
The evolution function $E_{_{NS}}(x,t)$, which describes the distribution
evolution from $t=0$ to $t$, is defined by
\begin{equation}
f_{_{NS}}(x,t) 
           = \int_x^1\frac{dy}{y} \, E_{_{NS}}(x/y,t) \, f_{_{NS}}(y,t=0) .
\label{eqn:evol}
\end{equation}
Then, it satisfies 
\begin{equation}
\frac{\partial}{\partial t} \, E_{_{NS}}(x,t) 
           = \int_x^1\frac{dy}{y} \, P_{_{NS}}(x/y) \, E_{_{NS}}(y,t) .
\label{eqn:evol2}
\end{equation}
Because this is the same integrodifferential equation as the original
DGLAP equation, one may wonder why such a function should be introduced.
There is an advantage that the evolution function should be
the delta function at $t=0$: $E_{_{NS}}(x,t=0)=\delta (1-x)$ because of
its definition in Eq. (\ref{eqn:evol}). It makes the following analysis
simpler. The functions are expanded in terms of the polynomials:
$P_{_{NS}}(e^{-x'}) = \sum_n P_{_{NS}}^n L_n (x')$ and
$E_{_{NS}}(e^{-x'},t) = \sum_n E_{_{NS}}^n (t) L_n (x')$,
where $P_{_{NS}}^n$ and $E_{_{NS}}^n (t)$ are the expansion coefficients.
The coefficient $F^n$ for a function $F(x)$ is given by 
$F^n=\int_0^1 dx L_n(x') \, F(x)$, and it could be calculated
analytically for a simple function.
If the two functions on the right-hand side of Eq. (\ref{eqn:evol2}) are
expanded, it becomes an integration of two Laguerre polynomials.
Using the formula
$\int_0^{x'} dy' L_n(x'-y') L_m(y') = L_{n+m}(x')-L_{n+m+1}(x')$
for this integration, we obtain
\begin{equation}
\frac{d}{d t} \, E_{_{NS}}^n (t) 
    = \sum_{m=0}^n (P_{_{NS}}^{n-m}-P_{_{NS}}^{n-m-1}) \, E_m (t).
\end{equation}
Because the evolution function is a delta function at $t=0$,
all the expansion coefficients are one. Therefore,
this equation is easily solved to give a summation form:
\begin{equation}
E_{_{NS}}^m (t) = e^{P_{_{NS}}^0 t}
    \sum_{k=0}^m \frac{t^k}{k!} \, B_m^k , \ \ \ 
B_m^{k+1}= \sum_{i=k}^{m-1} (P_{_{NS}}^{m-i}-P_{_{NS}}^{m-i-1}) B_i^k .
\end{equation}
This recursion relation is calculated with the relations
$B_i^0=1$, 
$B_i^1=\!\sum_{j=1}^i (P_{_{NS}}^j-P_{_{NS}}^{j-1})$, and
$B_0^k=B_1^k= \cdots =B_{k-1}^k=0$.
After all, the evolution is calculated by the simple summation:
\begin{equation}
f_{_{NS}}(x,t) = \sum_{n=0}^{N_{Lag}} \sum_{m=0}^{n} \,
    [ E_{n-m}(t) - E_{n-m-1}(t) ] \, L_n(-\ln \, x) f_{_{NS}}^{\ m} (t=0).
\end{equation}
In this way, the integrodifferential equation becomes a simple summation
of Laguerre-expansion coefficients, so that this method is considered
to be a very efficient numerical method for the solution.

\subsection{Mellin transformation method}
\label{mellin}

The Mellin transformation method is one of the popular evolution methods
\cite{mellin}. It is used because the Mellin transformation of the right-hand
side of Eq. (\ref{eqn:ns}) becomes a simple multiplication of two moments,
namely the moments of the splitting function and the distribution function.
The moments of the splitting functions (anomalous dimensions) are well
known, and a simple functional form is usually assumed for the distribution
at certain small $Q^2$ so as to calculate its moments easily.
Then, it is straightforward to obtain the analytical solution in
the moment space. Furthermore, the computation time is fairly short.
These are the reasons why this method has been
used as a popular method. For example, it is used for the $\chi^2$ analysis
of experimental data for obtaining polarized PDFs \cite{grsvbb},
whereas the brute-force method is employed in Ref. \cite{aac03}.

The Mellin transformation and inversion are defined by
\begin{equation}
\hat f(s,t)=\int_0^{1} dx \, x^{s-1} \, f(x,t) , \ \ \ \ 
f(x,t) =\frac{1}{2\pi i}\int_{c-i\infty}^{c+i\infty} ds \,
                    x^{-s} \, \hat f(s,t) .
\label{eqn:mellin}
\end{equation}
Here, the upper limit of the $x$ integration is taken one because
the distribution $f(x)$ vanishes in the region $x \ge 1$. The Mellin
inversion is a complex integral with an arbitrary real constant $c$,
which should be taken so that the integral $\int_0^1 dx f(x)x^{c-1}$ is
absolutely convergent. If this transformation is used,
the integrodifferential equations become very simple.
For example, the nonsinglet evolution equation becomes
\begin{equation}
\frac{\partial}{\partial t} \, \hat f_{_{NS}}(s,t)
               = \hat P_{_{NS}}(s) \, \hat f_{_{NS}}(s,t) .
\end{equation}
Its solution is simply given by
\begin{equation}
\hat f_{_{NS}}(s,t) = e^{\hat P_{_{NS}}(s) \, t}
                          \ \hat f_{_{NS}}(s,t=0) .
\label{eqn:ns1}
\end{equation}

\begin{wrapfigure}{r}{0.46\textwidth}
   \vspace{-0.3cm}
   \begin{center}
       \epsfig{file=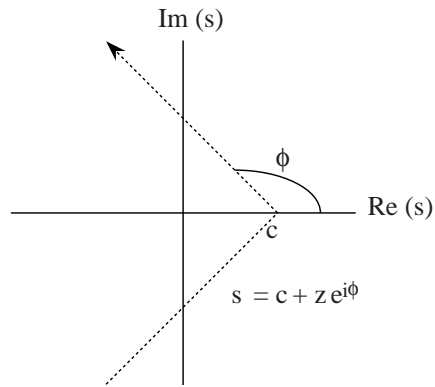,width=6.0cm}
   \end{center}
   \vspace{-0.5cm}
       \caption{\footnotesize Integration contour.}
   \vspace{+0.3cm}
       \label{fig:contour}
\end{wrapfigure}

Because the moments $\hat P_{_{NS}}(s)$ are well known quantities
and the moments of the initial function $\hat f_{_{NS}}(s,t=0)$
could be evaluated, it is straightforward to calculate the
evolution in Eq. (\ref{eqn:ns1}) in the moment space. However,
the numerical integration is needed for the Mellin inversion
in Eq. (\ref{eqn:mellin}) for transforming the moments into
a corresponding $x$ distribution.
Practically, the Mellin inversion is calculated along
the integration contour in Fig. \ref{fig:contour}. 
Changing the complex integration variable $s$ for the real one $z$
by $s=c+z e^{i \phi}$ in Eq. (\ref{eqn:mellin}), we have
\begin{equation}
f_{_{NS}}(x,t) = \frac{1}{\pi} \int_{0}^{\infty} dz \,
           Im \left[ e^{i \phi} \, x^{-c-ze^{i \phi}} \,
              \hat f_{_{NS}} (s=c+z e^{i \phi},t) \right] .
\label{eqn:mell-inv}
\end{equation}
The constant $c$ and the angle $\phi$ are shown in Fig. \ref{fig:contour}. 
Using the Gauss-Legendre quadrature for this integration, we 
obtain the evolved distribution in the $x$ space.

\section{Comparison of evolution results}
\label{results}

In comparing evolution results of three methods, we take the evolved
distribution of the brute-force (BF) method with $N_t$=200 and $N_x$=4000 as
a standard for assessing other evolution accuracy. It is shown
in Ref. \cite{bfsaga} that the evolution accuracy is better than 2\%
if $N_t$=200 and $N_x$=1000 are taken. This is the reason why it is
taken as the standard. Because the details are discussed in Ref.
\cite{bfsaga} for the evolution accuracy of the BF method,
we discuss only the comparison with the results of the Laguerre-polynomial
and Mellin-transformation methods.

\subsection{Parton distribution functions}
\label{pdfs}

In order to show the $Q^2$ evolution of the PDFs, we use the MRST02
distributions \cite{mrst02} which are provided analytically
at $Q^2$=1 GeV$^2$. The distributions are evolved to $Q^2$=100 GeV$^2$
with the MRST02 scale parameter by three evolution methods.
Then, the ratio of the evolved distribution to the one by
the brute-force method with $N_t$=200 and $N_x$=4000 is shown
for finding the numerical accuracy. 

\begin{figure}[h!]
\parbox[t]{0.46\textwidth}{
   \begin{center}
       \epsfig{file=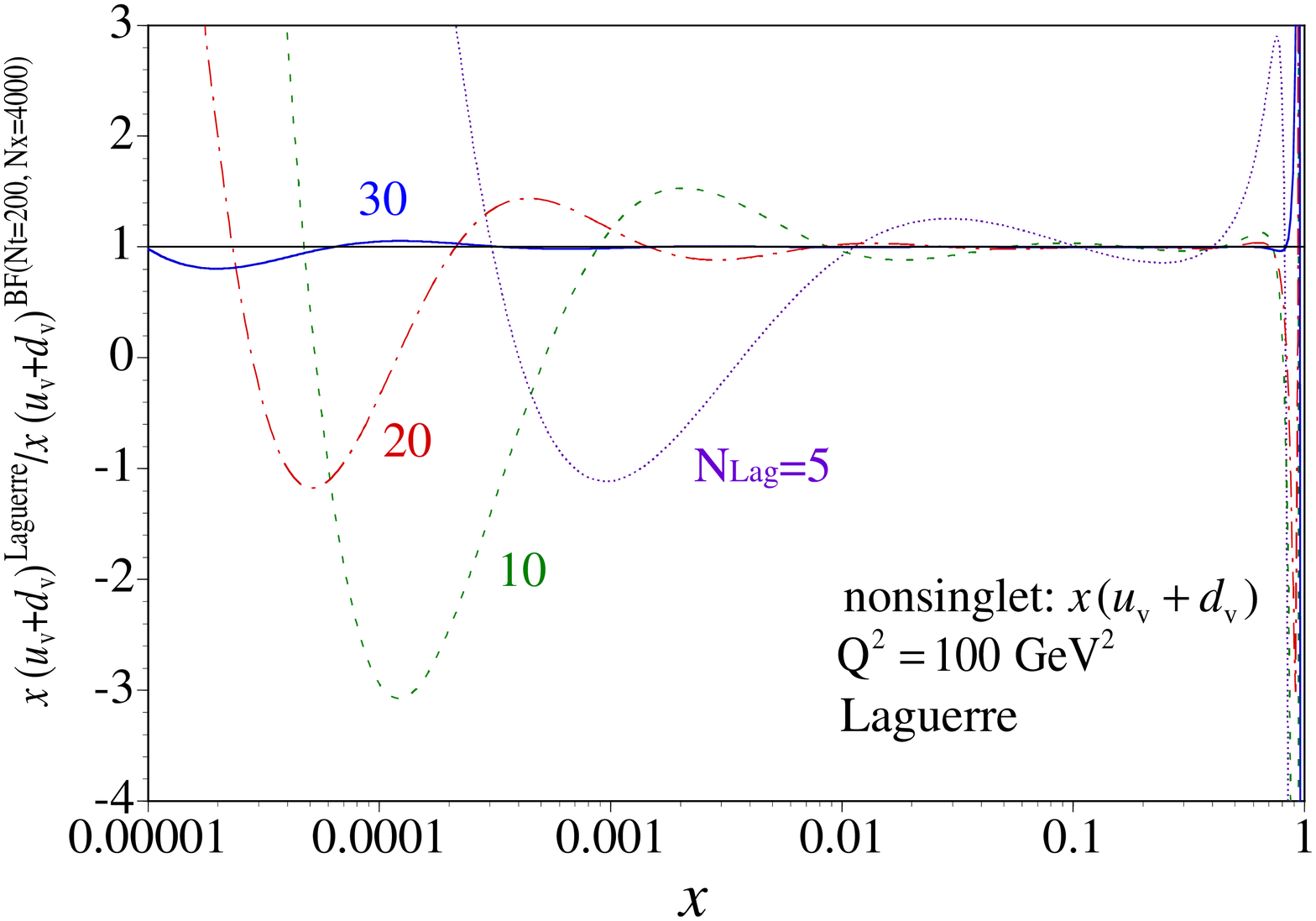,width=7.0cm}
   \end{center}
   \vspace{-1.4cm}
       \caption{\footnotesize
                Evolved nonsinglet distribution ratios 
                $x(u_v+d_v)^{Laguerre}/x(u_v+d_v)^{BF}$
                are shown for $N_{Lag}$=5, 10, 20, and 30 in
                the Laguerre-polynomial method.}
       \label{fig:pdf-ns-laguerre}
}\hfill
\parbox[t]{0.46\textwidth}{
   \begin{center}
   \epsfig{file=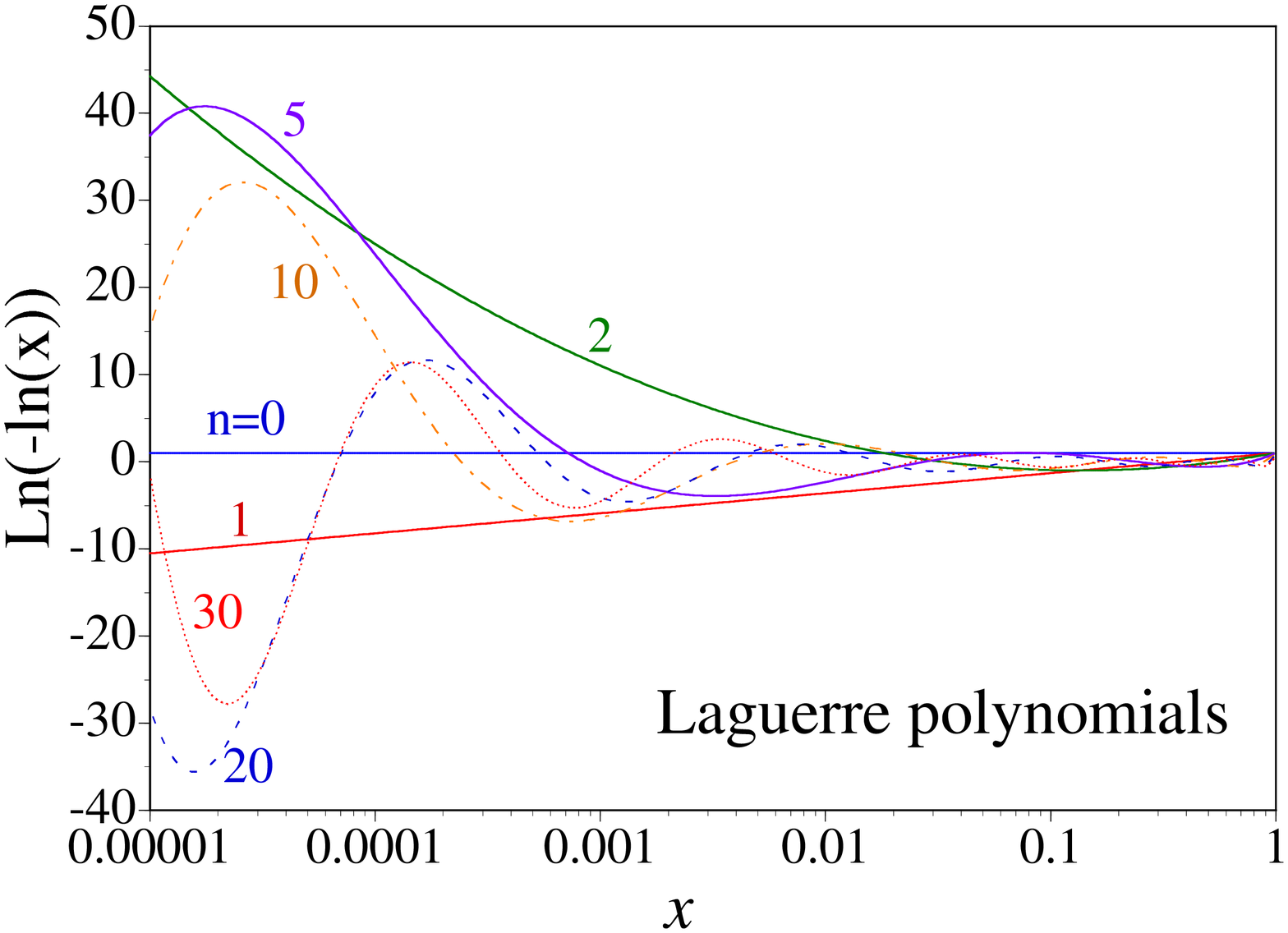,width=7.0cm}
   \end{center}
   \vspace{-1.4cm}
       \caption{\footnotesize
                Laguerre polynomials $L_n(-\ln x)$.}
       \label{fig:laguerre-poly}
}
\end{figure}

First, the evolution results of the nonsinglet distribution $x(u_v+d_v)$
are shown in Fig. \ref{fig:pdf-ns-laguerre} for the Laguerre method.
The number of the Laguerre polynomials $N_{Lag}$ is taken as
$N_{Lag}$=5, 10, 20, and 30, and each distribution ratio
$x(u_v+d_v)_{Laguerre}/x(u_v+d_v)_{BF(N_t=200,N_x=4000)}$ is shown.
It is obvious that accurate evolution cannot be obtained if the number
$N_{Lag}$ is small in the small- and large-$x$ regions.
In particular, the ratio shows oscillatory behavior at small $x$,
which results from the functional behavior of the Laguerre polynomials.
The Laguerre polynomials $L_n(-\ln \, x)$ are shown as a function of $x$
in Fig. \ref{fig:laguerre-poly}. We find that the oscillatory
functional form at small $x$ gives rise to the oscillatory behavior
in Fig. \ref{fig:pdf-ns-laguerre}. Therefore, one should be careful
in the Laguerre method that a large number of polynomials should be
taken to obtain an accurate evolution at small $x$. Furthermore,
one should be also careful in the very-large-$x$ region.

\vspace{-1.2cm}
\noindent
\vspace{-1.2cm}

\begin{figure}[h!]
\parbox[t]{0.46\textwidth}{
   \begin{center}
      \epsfig{file=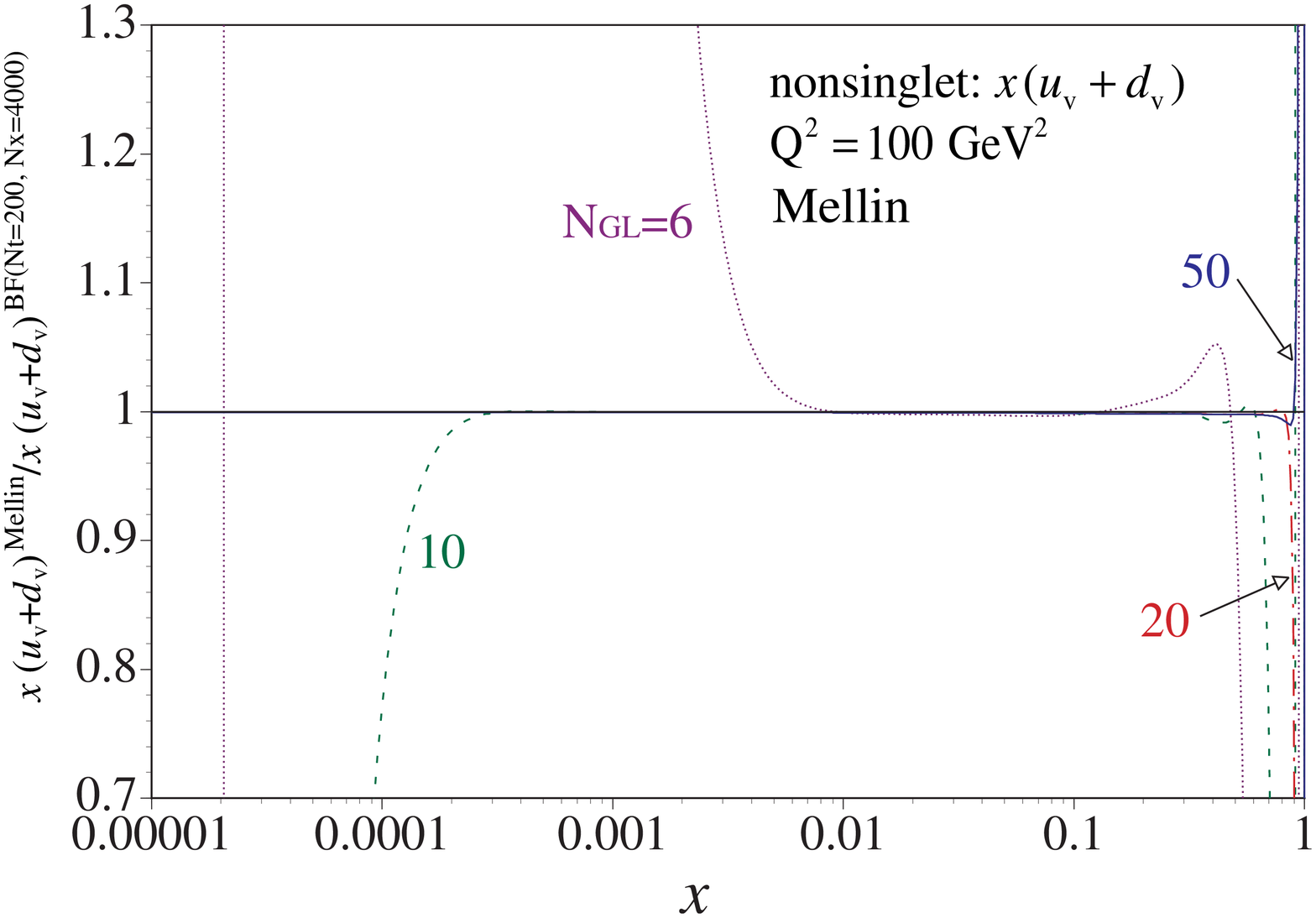,width=7.0cm}
   \end{center}
   \vspace{-1.4cm}
       \caption{\footnotesize
                Evolved nonsinglet distribution ratios 
                $x(u_v+d_v)^{Mellin}/x(u_v+d_v)^{BF}$
                are shown for $N_{GL}$=6, 10, 20, and 50 in
                the Mellin-transformation method.}
       \label{fig:pdf-ns-mellin}
}\hfill
\parbox[t]{0.46\textwidth}{
   \begin{center}
      \epsfig{file=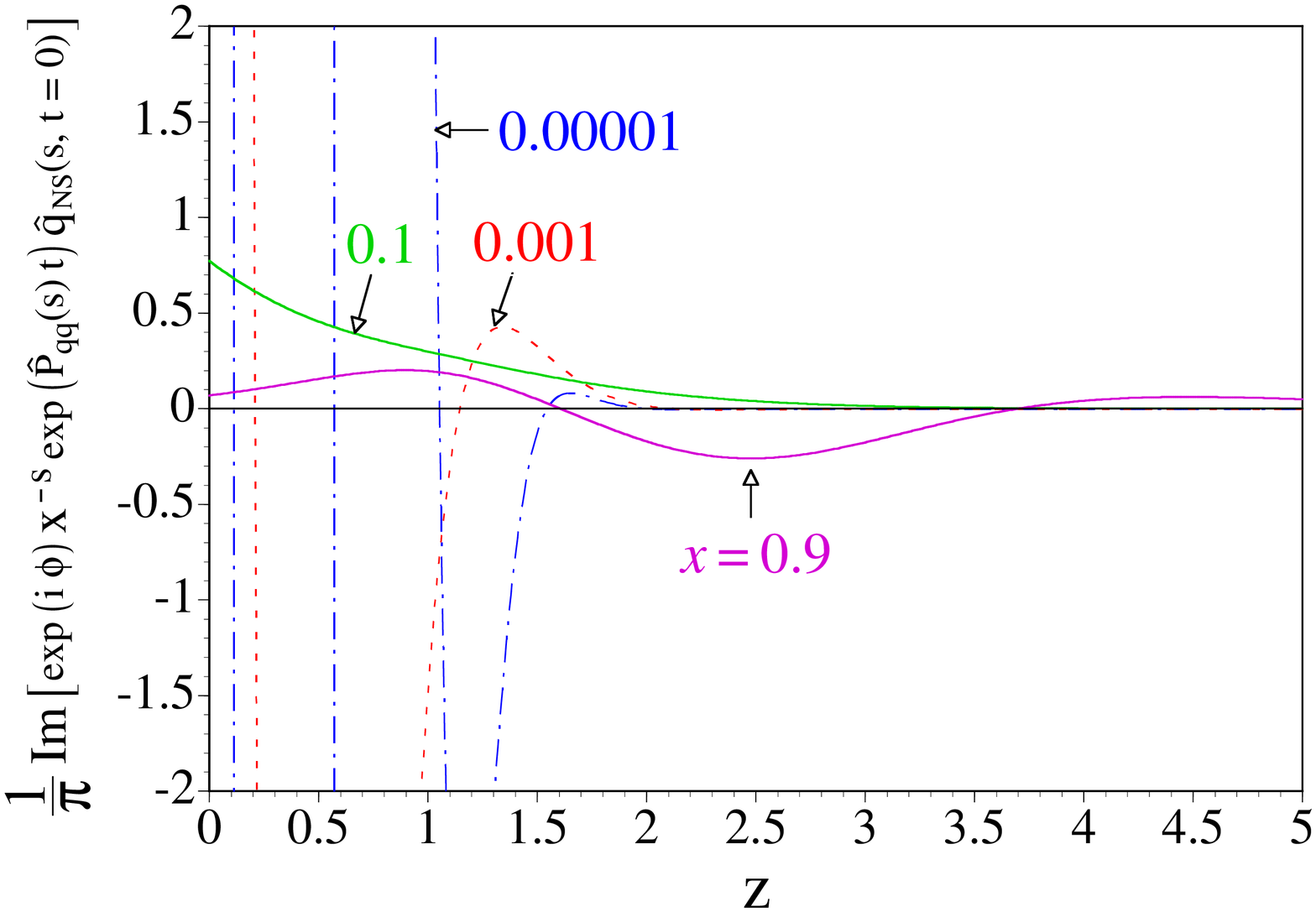,width=7.0cm}
   \end{center}
   \vspace{-1.4cm}
       \caption{\footnotesize
                Integrand of the Mellin inversion
                in Eq. (\ref{eqn:mell-inv}).}
        \label{fig:inv-integrand}
}
\end{figure}

Next, the evolution results are shown in Fig. \ref{fig:pdf-ns-mellin}
for the Mellin-transformation method. The Mellin inversion of
Eq. (\ref{eqn:mell-inv}) is numerically calculated by the Gauss-Legendre
quadrature with the number of points $N_{GL}$, which is taken as
$N_{GL}$=6, 10, 20, and 50 in Fig. \ref{fig:pdf-ns-mellin}.
The integration contour of Fig. \ref{fig:contour} is used with 
the constants $c$=1.1 and $\phi$=135$^\circ$ as suggested in
Ref. \cite{mellin}. In order to use the Gauss-Legendre quadrature,
the upper limit of the integration of Eq. (\ref{eqn:mell-inv}) should
be assigned. We decided to take $z_{max}=16.5x+3.5$ so that
the integrand is small enough at $z=z_{max}$ for
the $x$ region $10^{-5}<x<0.99$. If the number $N_{GL}$ is small,
the evolved nonsinglet distribution is not accurate enough in the small
and large $x$ regions as shown in Fig. \ref{fig:pdf-ns-mellin}.

\vspace{-1.4cm}
\noindent
\vspace{-0.5cm}

The inaccuracy in the small and large $x$ regions for $N_{GL}$=6 and 10
is understood in the following way. We show the integrand of
Eq. (\ref{eqn:mell-inv}) in Fig. \ref{fig:inv-integrand} by taking
$x$=$10^{-5}$, $10^{-3}$, $10^{-1}$, and 0.9. It is clear that
the integrand oscillates at small $x$
so that a certain number of Gauss-Legendre points is needed for getting
an accurate evolution. On the other hand, the integrand does not 
decrease rapidly at large $z$ for the large-$x$ case ($x$=0.9), 
so that large $z_{max}$ should be taken for the integration.
In addition, the positive contribution
at $z\sim 1$ and the negative one at $z\sim 2.5$ almost cancel each
other, which is another source of numerical inaccuracy.

\begin{figure}[h!]
\parbox[t]{0.46\textwidth}{
   \begin{center}
       \epsfig{file=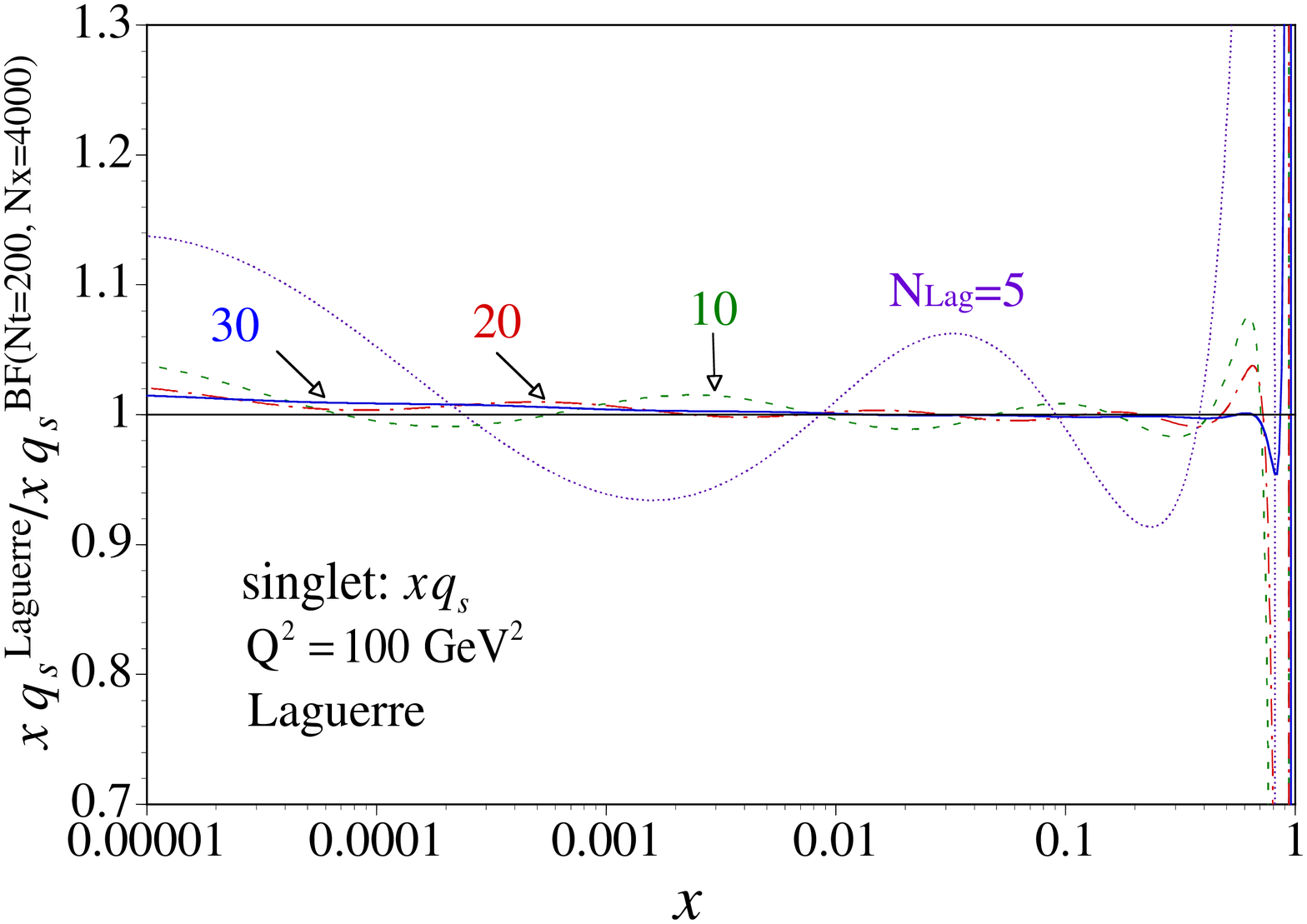,width=7.0cm}
   \end{center}
   \vspace{-1.4cm}
       \caption{\footnotesize
                Evolved singlet distribution ratios $xq_s^{Laguerre}/xq_s^{BF}$
                are shown for $N_{Lag}$=5, 10, 20, and 30 in
                the Laguerre-polynomial method.}
       \label{fig:pdf-qs-laguerre}
}\hfill
\parbox[t]{0.46\textwidth}{
   \begin{center}
   \epsfig{file=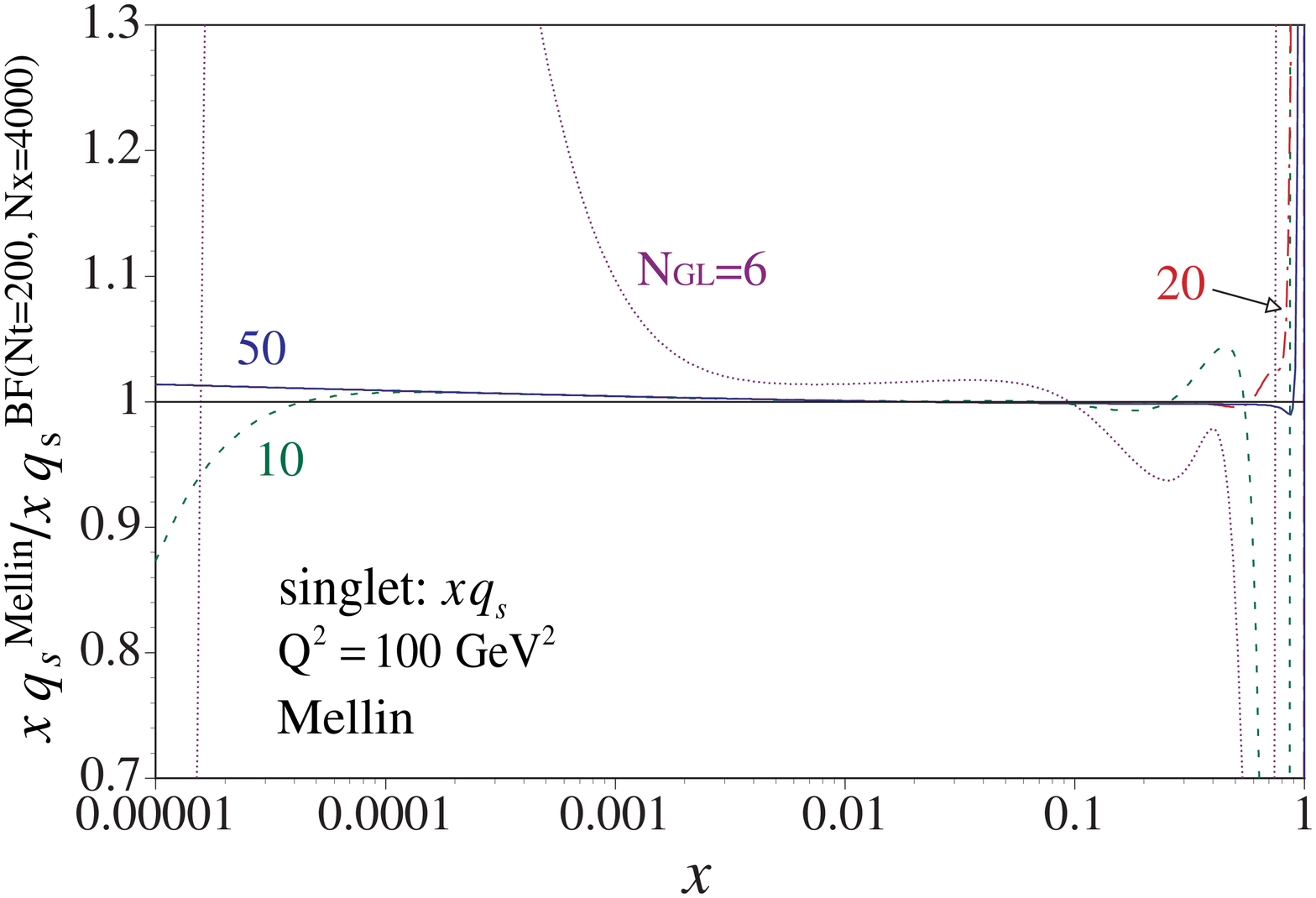,width=7.0cm}
   \end{center}
   \vspace{-1.4cm}
       \caption{\footnotesize
                Evolved singlet distribution ratios $xq_s^{Mellin}/xq_s^{BF}$
                are shown for $N_{GL}$=6, 10, 20, and 50 in
                the Mellin-transformation method.}
       \label{fig:pdf-qs-mellin}
}
\end{figure}

The evolution results of the singlet distribution are shown in 
Figs. \ref{fig:pdf-qs-laguerre} and \ref{fig:pdf-qs-mellin}
for the Laguerre and Mellin methods, respectively.
We notice in Fig. \ref{fig:pdf-qs-laguerre} that the accuracy of
the singlet evolution is much better than the nonsinglet one in
the Laguerre method. It comes simply from the functional difference
between the nonsinglet and singlet distributions. The Laguerre method
can be used as an accurate evolution method for the singlet evolution.
The singlet evolution accuracy for the Mellin method is similar to
its nonsinglet ones. If the point number $N_{GL}$ is large enough,
the evolution becomes accurate.

\vspace{-1.5cm}
\noindent
\vspace{-0.5cm}

\begin{figure}[h!]
\parbox[t]{0.46\textwidth}{
   \begin{center}
       \epsfig{file=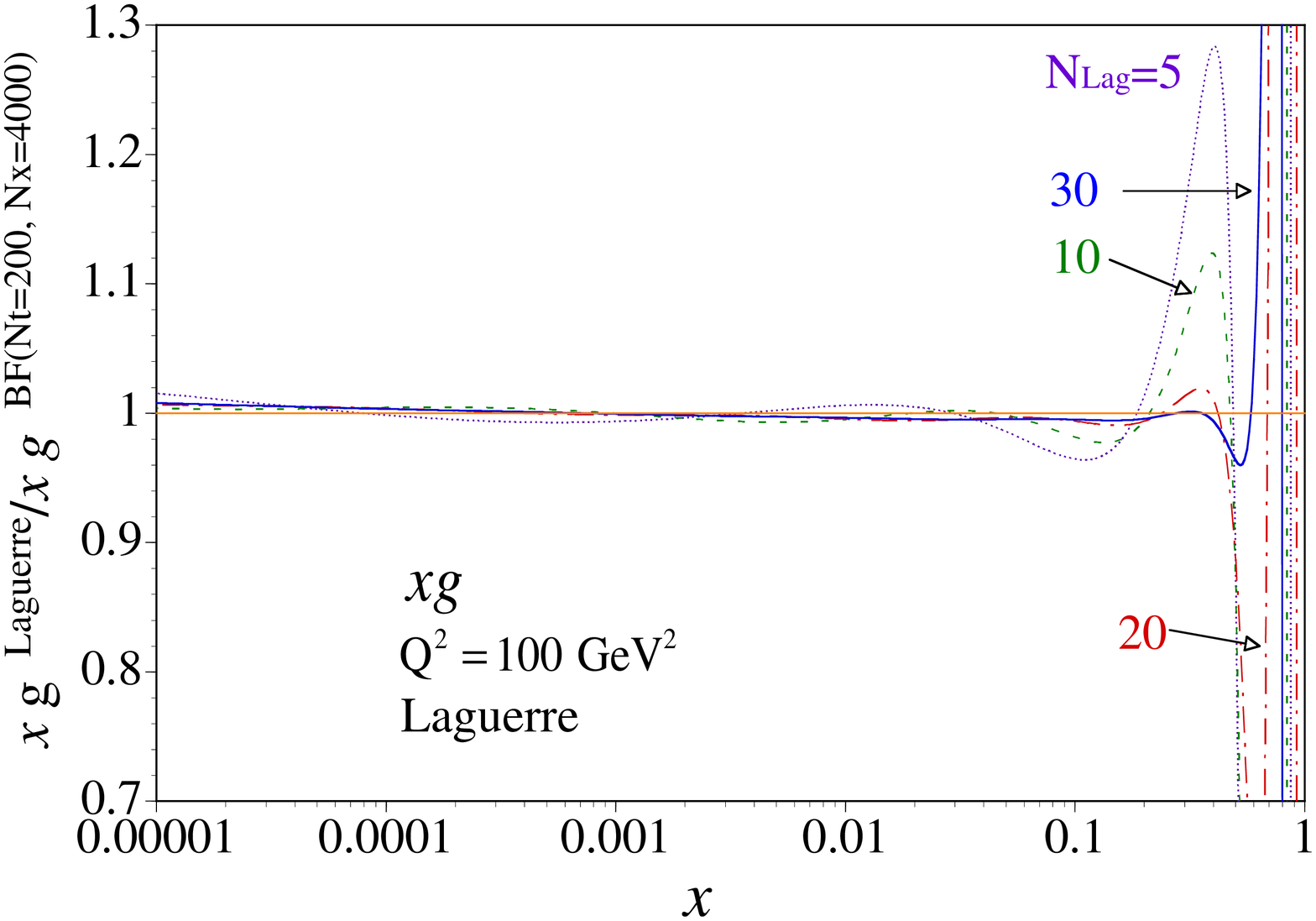,width=7.0cm}
   \end{center}
   \vspace{-1.4cm}
       \caption{\footnotesize
                Evolved gluon distribution ratios $xg^{Laguerre}/xg^{BF}$
                are shown for $N_{Lag}$=5, 10, 20, and 30 in
                the Laguerre-polynomial method.}
       \label{fig:pdf-g-laguerre}
}\hfill
\parbox[t]{0.46\textwidth}{
   \begin{center}
   \epsfig{file=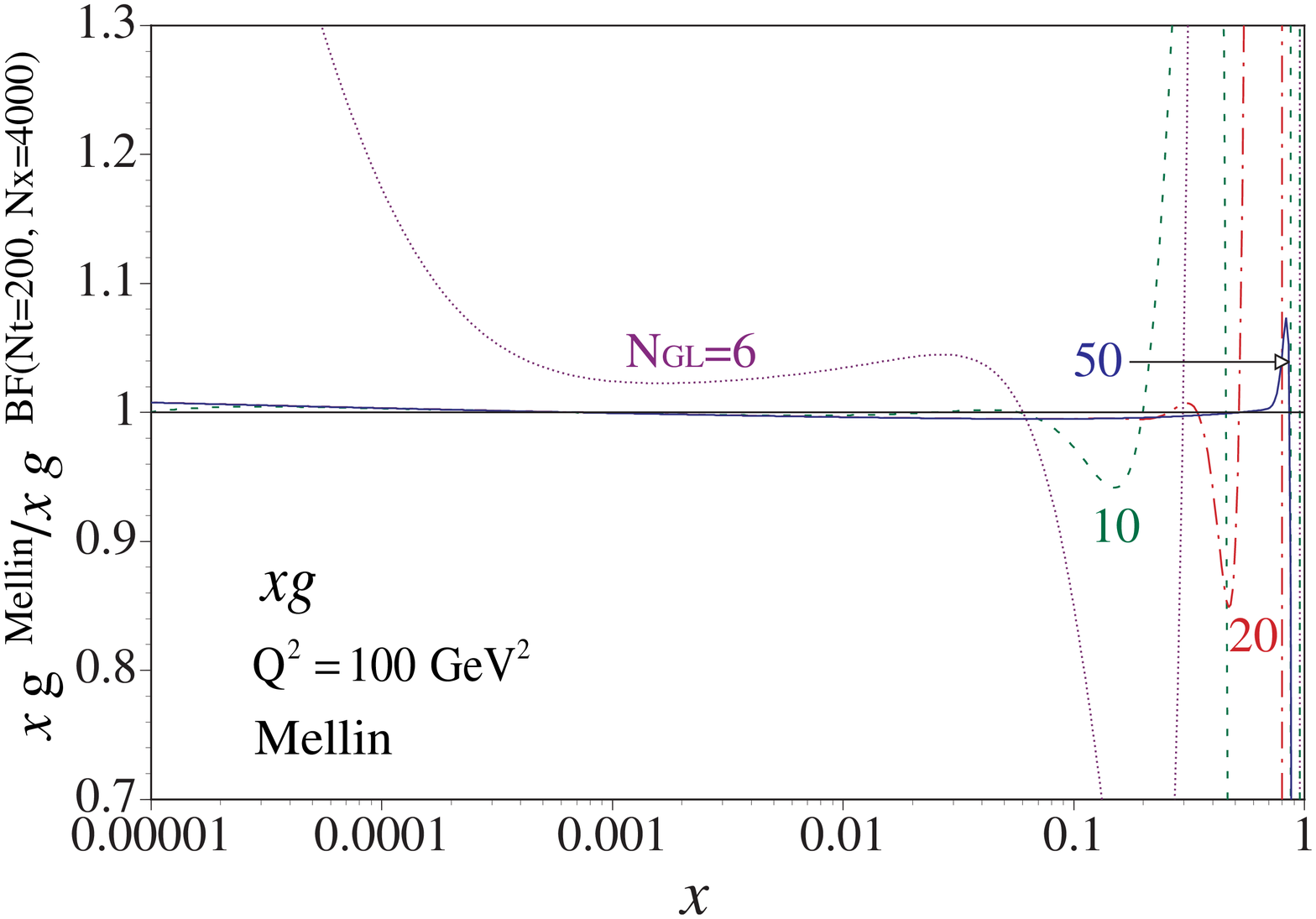,width=7.0cm}
   \end{center}
   \vspace{-1.4cm}
       \caption{\footnotesize
                Evolved gluon distribution ratios $xg^{Mellin}/xg^{BF}$
                are shown for $N_{GL}$=6, 10, 20, and 50 in
                the Mellin-transformation method.}
       \label{fig:pdf-g-mellin}
}
\end{figure}

The gluon evolution results are shown in Figs. \ref{fig:pdf-g-laguerre}
and \ref{fig:pdf-g-mellin} for the Laguerre and Mellin methods.
The Laguerre evolution becomes much more accurate than its singlet-quark
evolution; however, the Mellin evolution becomes slightly inaccurate
at large $x$. 

\vspace{-1.5cm}
\noindent
\vspace{-0.7cm}

\subsection{Fragmentation functions}
\label{ffs}

The fragmentation functions (FFs) are essential for understanding hadron
productions in high-energy reactions. In addition, they are important
for describing ultrahigh-energy cosmic rays in the top-down scenario
\cite{toldra,topdown}.
In comparison with the situation of the PDFs, their determination is still
premature in the sense that experimental data are not still enough to determine
them accurately. However, there are available $e^+ e^-$ annihilation data,
which could be used for a global analysis of the FFs. The current status
of such analyses is summarized in Ref. \cite{ffs}. We use the KKP
parametrization \cite{kkp} at $Q^2$=2 GeV$^2$ as the initial functions,
and then they are evolved to $Q^2$=100 GeV$^2$ with the KKP scale parameter
for testing three evolution methods.

\begin{figure}[h]
\parbox[t]{0.46\textwidth}{
   \begin{center}
       \epsfig{file=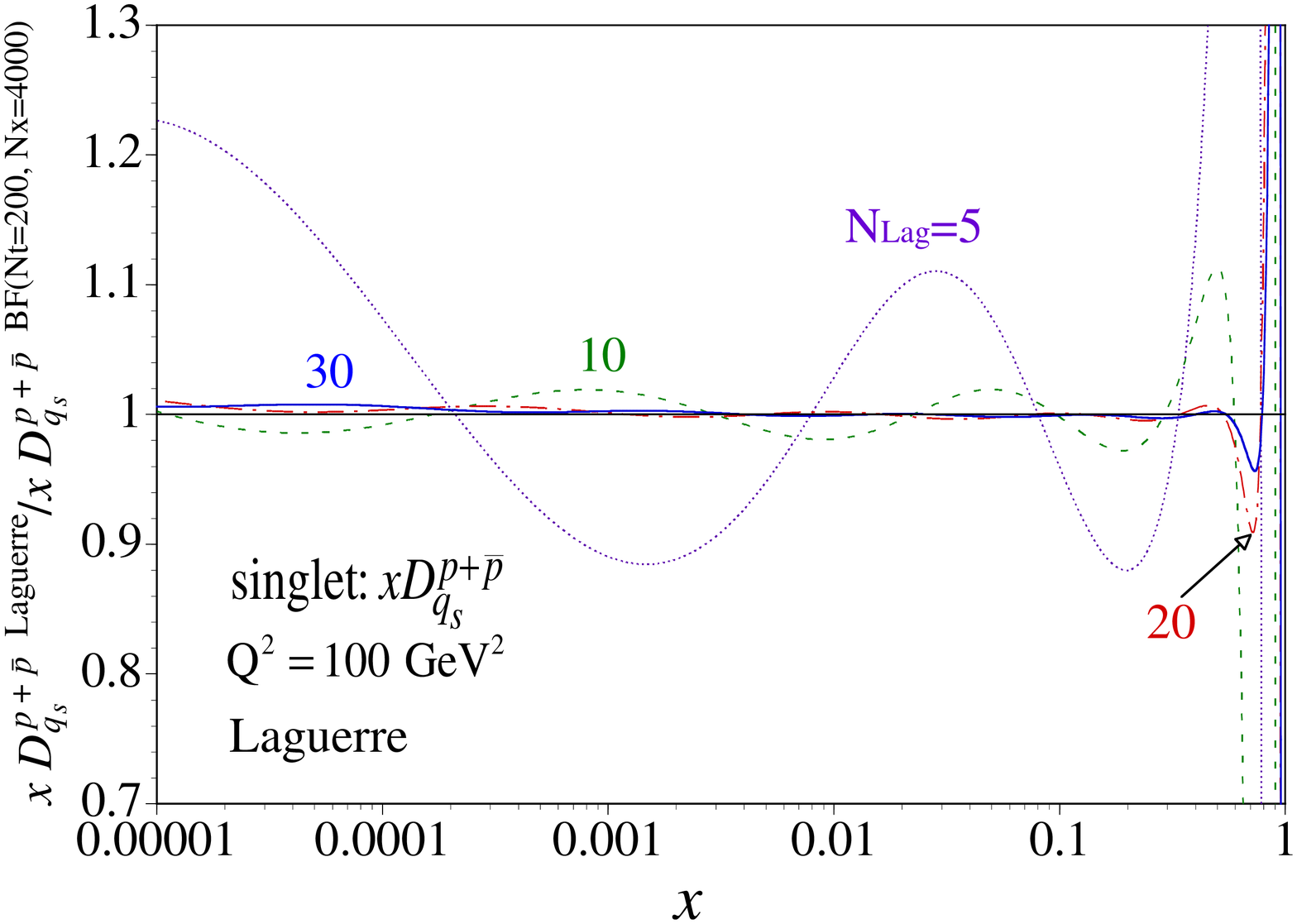,width=7.0cm}
   \end{center}
   \vspace{-1.4cm}
       \caption{\footnotesize
                Evolved singlet fragmentation function ratios
                $xD_{q_s}^{p+\bar p, \ Laguerre}/xD_{q_s}^{p+\bar p, \ BF}$
                are shown for $N_{Lag}$=5, 10, 20, and 30 in
                the Laguerre-polynomial method.}
       \label{fig:ff-qs-laguerre}
}\hfill
\parbox[t]{0.46\textwidth}{
   \begin{center}
   \epsfig{file=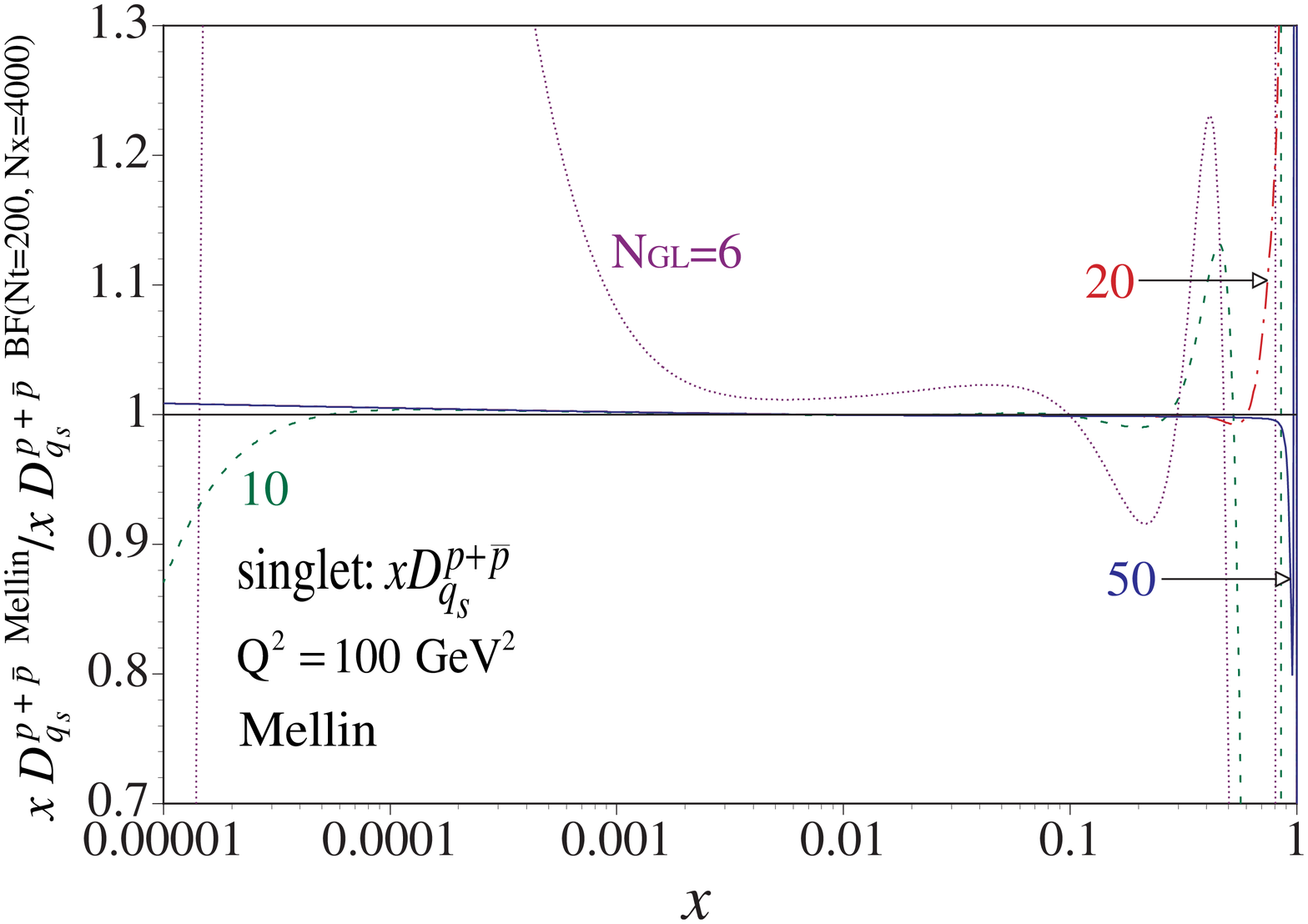,width=7.0cm}
   \end{center}
   \vspace{-1.4cm}
       \caption{\footnotesize
                Evolved singlet fragmentation function ratios
                $xD_{q_s}^{p+\bar p, \ Mellin}/xD_{q_s}^{p+\bar p, \ BF}$
                are shown for $N_{GL}$=6, 10, 20, and 50 in
                the Mellin-transformation method.}
       \label{fig:ff-qs-mellin}
}
\end{figure}

The brute-force evolution with $N_t$=200 and $N_x$=4000 
is taken as the standard for showing other evolution results as it was
done in the previous subsection. In Fig. \ref{fig:ff-qs-laguerre},
the Laguerre-method results for the singlet fragmentation function
into the proton and antiproton, 
$D_{q_s}^{p+\bar p}=\sum_i (D_{q_i}^{p+\bar p}+D_{\bar q_i}^{p+\bar p})$,
are shown. The Mellin method results are shown in Fig. \ref{fig:ff-qs-mellin}.
The small-$x$ region is shown in these figures for comparing 
the results with the PDF accuracy in Sec. \ref{pdfs},
although it is outside the range of current accelerator experiments.
As it was found in the PDF evolution, the Laguerre method is not
excellent in the small- and large-$x$ regions unless
a large number of polynomials is taken. The ratios in 
Figs. \ref{fig:ff-qs-laguerre} and \ref{fig:ff-qs-mellin}
show a similar tendency to the ratios of the PDF singlet evolution 
results in Figs. \ref{fig:pdf-qs-laguerre} and \ref{fig:pdf-qs-mellin},
respectively.

\vspace{-1.0cm}
\noindent
\vspace{-0.9cm}

\vspace{-0.8cm}
\noindent
\begin{figure}[h]
\parbox[t]{0.46\textwidth}{
   \begin{center}
       \epsfig{file=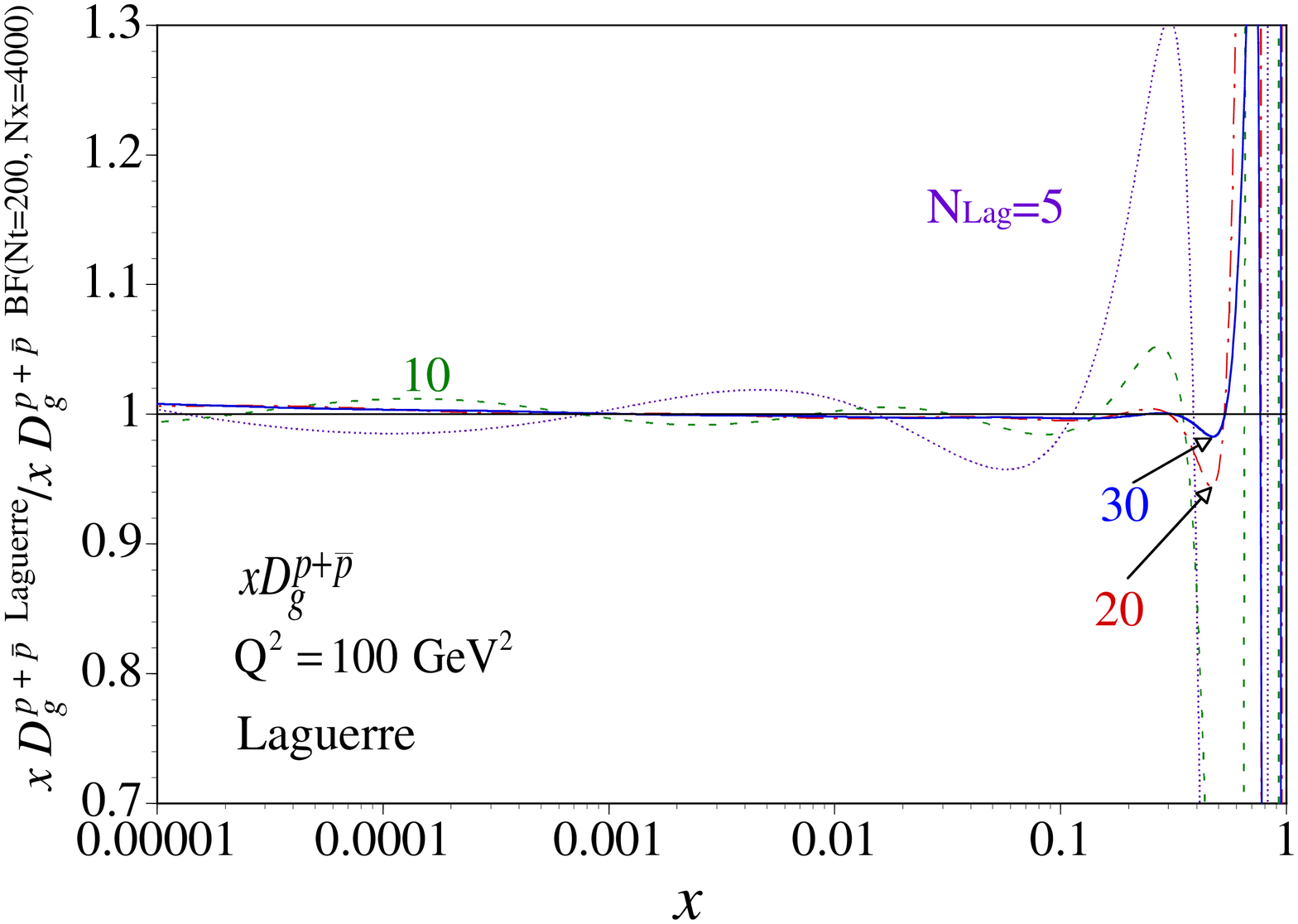,width=7.0cm}
   \end{center}
   \vspace{-1.4cm}
       \caption{\footnotesize
                Evolved gluon fragmentation function ratios
                $xD_{g}^{p+\bar p, \ Laguerre}/xD_{g}^{p+\bar p, \ BF}$
                are shown for $N_{Lag}$=5, 10, 20, and 30 in
                the Laguerre-polynomial method.}
       \label{fig:ff-g-laguerre}
}\hfill
\parbox[t]{0.46\textwidth}{
   \begin{center}
   \epsfig{file=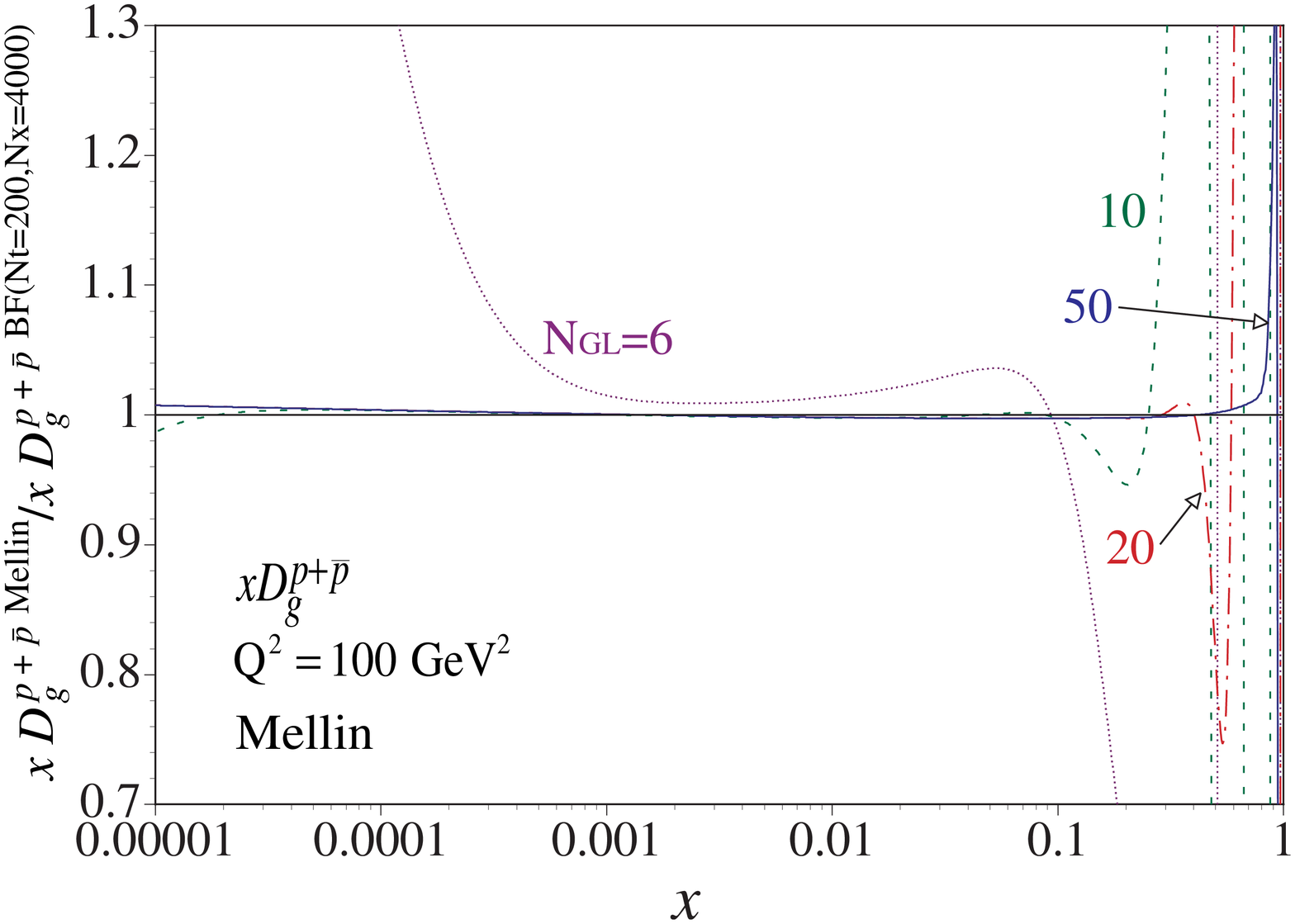,width=7.0cm}
   \end{center}
   \vspace{-1.4cm}
       \caption{\footnotesize
                Evolved gluon fragmentation function ratios
                $xD_{g}^{p+\bar p, \ Mellin}/xD_{g}^{p+\bar p, \ BF}$
                are shown for $N_{GL}$=6, 10, 20, and 50 in
                the Mellin-transformation method.}
       \label{fig:ff-g-mellin}
}
\end{figure}
\vspace{-0.5cm}

The gluon FF evolution results are shown in Figs. \ref{fig:ff-g-laguerre}
and \ref{fig:ff-g-mellin} for the Laguerre and Mellin methods.
The gluon FF evolution by the Laguerre method is accurate
in the most-$x$ region except for the large-$x$ part.  
The Mellin method is also accurate except for the large-$x$ region. 

We have compared three evolution methods; however, there are other 
methods \cite{others}. Because the numerical solution of
the DGLAP equations is very important for describing high-energy
hadron reactions, an efficient and accurate method should be investigated
further.

\subsection{Computation time}
\label{time}

We show typical computation time for each evolution method.
However, we should aware that it depends much on the numbers,
$N_t$ and $N_x$ in the brute-force method, $N_{Lag}$ in the Laguerre method,
and $N_{GL}$ in the Mellin method. Therefore, we list CPU time for different 
parameter values for $N_t$, $N_x$, $N_{Lag}$, and $N_{GL}$ in
Table \ref{tab:cpu} by running the codes for the nonsinglet PDF.
In the singlet evolution, the time becomes longer but the tendency
of three evolution methods is the same.
The used machine is DELL-Dimension-8800 with a Pentiam-4 2.8G CPU.
The operating system is Redhat-Linux 8.0 and the fortran complier is g77.

\vspace{-0.4cm}
\begin{table}[h]
\caption{CPU time for calculating PDFs at 500 $x$-points
         by running each evolution code for the nonsinglet
         distribution with the linux-g77 compiler
         on a Pentiam-4 machine with a 2.8G CPU.}
\label{tab:cpu}
\vspace{0.2cm}
\begin{center}
\begin{tabular}{|c|c|c|}
\hline
Method       &  Condition              & CPU time (seconds)      \\
             &                         & for PDFs at 500 $x$-points \\
\hline\hline
Brute-force  & $N_t$=50, $N_x$ =1000   &    1.501                \\
\cline{2-3}
             & $N_t$=200, $N_x$ =1000  &    5.986                \\
\cline{2-3}
             & $N_t$=200, $N_x$ =4000  &   95.634   \            \\
\hline\hline
Laguerre     & $N_{Lag}$=5             &    0.005                \\   
\cline{2-3}
             & $N_{Lag}$=10            &    0.011                \\   
\cline{2-3}
             & $N_{Lag}$=20            &    0.025                \\   
\cline{2-3}
             & $N_{Lag}$=30            &    0.044                \\   
\hline\hline
Mellin       & $N_{GL}$=6              &    0.154                \\   
\cline{2-3}
             & $N_{GL}$=10             &    0.244                \\   
\cline{2-3}
             & $N_{GL}$=20             &    0.464                \\   
\cline{2-3}
             & $N_{GL}$=50             &    1.128                \\   
\hline
\end{tabular}
\end{center}
\end{table}
\vspace{-0.8cm}

Among the three methods, the brute-force method takes the longest time
for the computation simply because the large numbers
of steps are taken. Therefore, it typically takes a few seconds for
obtaining a reasonable accuracy for the nonsinglet evolution
($N_t$=50-200, $N_x$=1000).

In the Laguerre method, the evolution is simply given by the summation
of the Laguerre coefficients which are calculated partially with
the recursion relation. There is no numerical integration involved
in the evolution calculation, so that this method is by far the fastest
among the studied methods. Even if $N_{Lag}$=30 is taken, it takes
0.044 seconds for the nonsinglet evolution. It means that it is 
one hundred times faster than the brute-force method. If one is interested
in using it for the singlet evolution and if one does not mind one
or two percent error, it is certainly the best method. 

In the Mellin method, accurate evolution results are obtained with
$N_{GL}=20$. The computation time is significantly shorter and
it is several times faster than the brute-force method.
This is the reason why this method is popular among high-energy physics
researchers.

\section{Summary}
\label{summary}

We have compared the evolution results of the parton distribution functions
and the fragmentation functions by using three evolution methods,
brute-force, Laguerre-polynomial, Mellin-transformation methods.
The advantages and disadvantages of each method are summarized in
Table \ref{tab:adv-dis}.

\vspace{-0.4cm}
\begin{table}[h]
\caption{Summary of advantages and disadvantages of each evolution method.}
\label{tab:adv-dis}
\vspace{0.2cm}
\begin{center}
\begin{tabular}{|l||p{6.5cm}|p{6.5cm}|}
\hline
Method     & Advantage       & Disadvantage                      \\
\hline\hline
Brute-force & Simple code: 
             The computer code is very simple. More complicated evolution
             equations with higher-twists ({\it e.g.} in Ref. \cite{mq})
             could be handled easily. The evolution could be accurate 
             in the small- and large-$x$ regions.
           & Long computation time:
             In order to obtain an accurate evolution, large numbers
             of steps ($N_t$ and $N_x$) are needed. If one uses the code
             for many evolution calculations, it takes a significant amount
             of time.
             \\
\hline
Laguerre   & Very fast: It takes less than a second by an ordinary
             desktop computer. As long as one does not mind the very small-
             and large-$x$ regions, it is a good method for
             repeated evolution calculations.
           & Accuracy at small and large $x$: 
             Depending on the initial functional form, the results do not
             converge at small $x$ unless a large number of polynomials
             are used. It is also difficult to obtain accurate evolution
             at large $x$.
             \\
\hline
Mellin     & Fast: By choosing an appropriate $z_{max}$ and $N_{GL}$
                   in each $x$ region, the code becomes much faster than
                   the brute-force computation. For repeated evolution
                   calculations with certain accuracy, this method
                   is appropriate.
           & Accuracy at small and large $x$: One should be careful about
             the choices of $z_{max}$ and $N_{GL}$. 
             In particular, the Mellin inversion should be carefully done 
             at very large $x$.
             \\
\hline
\end{tabular}
\end{center}
\end{table}
\vspace{-0.8cm}

\vfill\eject

\section*{Acknowledgments}
S.K. was supported by the Grant-in-Aid for Scientific Research from
the Japanese Ministry of Education, Culture, Sports, Science, and Technology. 
He also thanks the Institute for Nuclear Theory at the University of
Washington for its hospitality and the Department of Energy for partial
support during the completing of this work. 


\end{document}